## Ars combinatoria y modelos cosmológicos: correspondencias y pervivencias en una pintura cuzqueña del siglo XVIII [1]

Alejandro Gangui / Argentina
Gabriela Siracusano / Argentina
Juan Ricardo Rey Márquez / Argentina

> Siete son los grados de la contemplación: el primero es el fuego; el segundo es la unción; el tercero es el éxtasis; el cuarto, la contemplación; el quinto, el gusto; el sexto, el descanso, el séptimo, la gloria; después de lo cual no queda más que la felicidad eterna[2]

### Introducción

«Hízome el Señor una merced muy grande en una visión imaginaria» escribió María de Ágreda en el siglo XVII, «púsome su Majestad al pie de una bellísima Escala, y mostróme había de subir por ella». Estas palabras remiten al ascenso espiritual, presente en la tradición judeo-cristiana, cristalizada en las visiones de profetas y santos católicos. En el Génesis (18: 10-22) se narra que Jacob, yendo hacia Jarán, se durmió sobre unas piedras y vio una escala entre el cielo y la tierra por la cual ascendían y descendían ángeles; en el sueño Dios le prometió la tierra a sus descendientes –los israelitas– por lo que al despertarse Jacob señaló el lugar con una estela de piedras ungidas en aceite y la nombró Betel que significa «la casa de Dios»[3]. A partir de este sueño, la unión simbólica entre cielo y tierra se ha figurado con la escalera, adjudicándole diversas significaciones con los siglos. En los grados de la escalera se vio una metáfora de la gradualidad del ascenso; Benedicto (c. 480-547) usó una escala de doce pasos de humildad en su *Regula*, y en el siglo VII d.C. Juan Clímaco, Obispo de Sinaí, estableció una *Scala Paradisi* de treinta pasos de meditación que conducían a Dios[4]. Este modelo se mantuvo durante siglos, como lo muestran las cuatro centurias que separan el epígrafe de san Buenaventura de la visión de María de Ágreda. En cada evocación del símbolo se actualizó su sentido, sobre todo en lo relacionado con la imagen del cielo, pues si bien la escalera es una metáfora efectiva y directa, aquel -en cambio- es un lugar sagrado, visible sólo en parte pero al que había que conducir a los fieles *ad invisibilia per visibilia* según San Pablo[5]. En el presente texto trataremos los ecos históricos de la regla benedictina, la voz de Clímaco en el desierto y la visualización de los misterios celestes, en *Alegoría del firmamento con la representación de los siete cielos de los planetas* (siglo XVIII). Con este objetivo pretendemos analizar el cruce entre ciencia y mística en la representación de las escaleras celestes, para mostrar el anclaje cultural de la simbología utilizada en estos casos.

[1] Los autores agradecen el apoyo financiero del Conicet de la República Argentina, de la Universidad de Buenos Aires, de la Universidad Nacional de San Martín y de la ANPCyT, proyecto Pict 2011-1327, imprescindible para la realización de este estudio.
[2] San Buenaventura. Commentarius in Evangelium Sancti Lucae, c. 1248-1250.
[3] Hall, [1974] 1996, p. 210.
[4] Demaray, 1987, p. 28.

### 1. El debate sobre el cosmos

La visión de Jacob fue representada en íconos y manuscritos como una escalera que conduce a las nubes, a la morada divina o a la propia presencia de Dios, en una visión dual del universo compuesto por tierra y cielo. Pero esta visión se transformó con el redescubrimiento del *De Caelo* (350 a.C.) de Aristóteles por vía musulmana. Desde entonces la Creación se concibió como un todo coherente, dividido en el mundo terrenal (sublunar), el astral (supralunar) y el empíreo. En el mundo sublunar viven los cuatro elementos de Empédocles, dispuestos, desde el centro de la Tierra y hacia arriba, en el orden tierra, agua, aire y fuego. Aristóteles afirma que todo elemento tiende a dirigirse hacia su lugar natural, lo más rápidamente posible y por el camino más corto. El lugar natural de los objetos pesados es el centro de la Tierra y hacia allí caen en línea recta. Por el contrario, el fuego asciende, también en línea recta, tratando de alcanzar el límite máximo de la región terrestre, la *esfera* lunar.

La región supralunar albergaba a los astros, y para describirlos Aristóteles adoptó el modelo platónico de las esferas, ubicando las estrellas fijas equidistantes al centro de la Tierra y explicando que más allá de su esfera, no había espacio ni tiempo. Sólo una cualidad sobrehumana, un «dios aristotélico», estaría después de la última esfera imprimiéndoles el movimiento de rotación observado. Este *Primum Mobile* (o Primer Motor), según su doctrina, ha impulsado al mundo desde siempre y lo hará por siempre, pues dada su naturaleza móvil, sería absurdo pensar que permaneciera quieto por un tiempo infinito hasta el instante en que comenzó a trabajar. Para Aristóteles, la sustancia primordial de la región supralunar es ese elemento incorruptible e imponderable de Platón, el *éter*, cuya esencia cristalina y transparente se condensa para formar los astros. En su visión, el cosmos es eterno: no ha sido creado ni tendrá un final. La sustancia original de los cielos los convierte en eternamente iguales a sí mismos, y la naturaleza del éter los mantiene en continuo movimiento, el más perfecto, aquel en el que no existe comienzo ni final, el movimiento circular.

[5] Duby, 2011, p. 10.

La recuperación del pensamiento griego en la Europa Medieval lentamente excedió los muros de los monasterios y escuelas religiosas, desde donde se difundió el estudio de los autores clásicos. Pero algunas ideas aristotélicas iban en contra de las Sagradas Escrituras. La proposición de un mundo eterno chocaba de lleno con la doctrina bíblica de la creación del Génesis. Aristóteles había establecido, entre otras cosas, la inexistencia del vacío y la finitud del espacio, con lo que se ponía en cuestión la omnipotencia divina. Aristóteles es entonces criticado y cuestionado, prohibido incluso en universidades como la de París en 1210. Sin embargo, la visión griega seduce a los intelectuales que buscaron sintetizar la razón de los pensadores griegos y la fe cristiana, obra que fue llevada a cabo por el teólogo y filósofo cristiano Tomás de Aquino. Santo Tomás unió las enseñanzas del Estagirita con las doctrinas de la Iglesia, formando un sistema de pensamiento que duraría siglos[6].

Aun así, una importante cuestión a resolver era hacer encajar a Dios en el rígido sistema aristotélico. Anselmo, arzobispo de Canterbury, propuso como adecuada morada divina al *Empíreo*, el éter luminoso *abrasado*, bautizado así por el neoplatónico Proclo. Para la teología medieval, el Empíreo era una esfera de fuego exterior a la de las estrellas fijas, creada en el primer día del Génesis, en donde se hallaría Dios. A mediados del siglo XIV, el teólogo y matemático de Oxford Thomas Bradwardine, extenderá el empíreo propuesto por Anselmo a un vacío extramundano infinito. Pues en su visión, más allá de la esfera de las estrellas fijas se extendía el reino sin límites de Dios[7].

El modelo geocéntrico del universo consagrado en el *Almagesto*, obra cumbre del astrónomo griego Claudio Ptolomeo (siglo II d.C.), fue aceptado y adaptado desde ese momento por la filosofía y las religiones cristiana y musulmana. Aunque en 1543 se introduce la teoría heliocéntrica de Copérnico y un siglo más tarde las revolucionarias ideas de Galileo habían ya recorrido buena parte del mundo conocido, la concepción cosmológica medieval todavía primaba en centros de estudio y en talleres de artesanos y artistas. Esto puede verse en la representación de las esferas de varios grabados, entre los que se destacan dos contemporáneos a Copérnico: las *Crónicas de Nüremberg* (1493) de Hartmann Schedel, y la *Cosmographia* de Pedro Apiano (1524) (Fig. 1).

---

[6] Duby, 2011, p. 78.
[7] Lindberg, 1992.

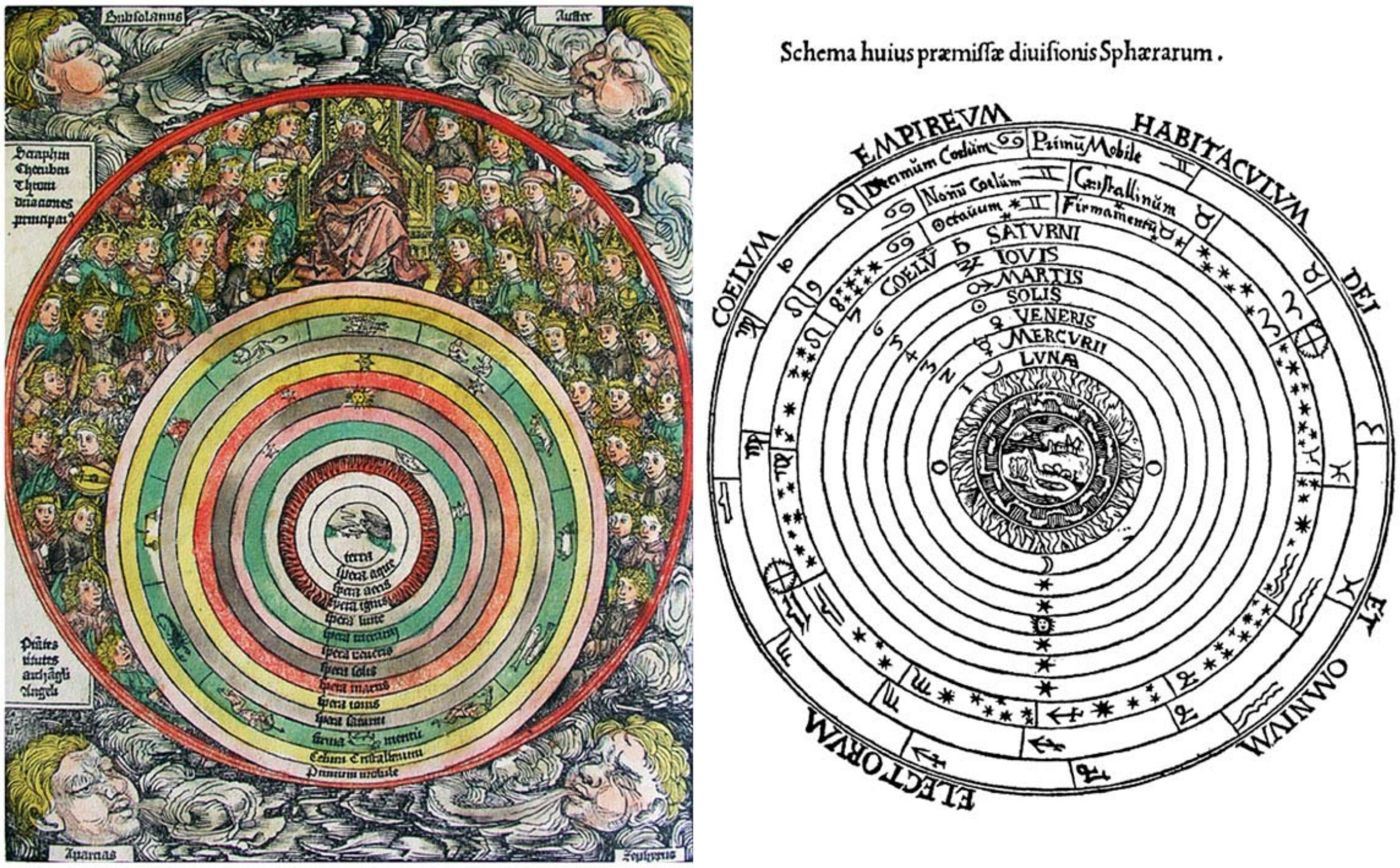

Figura 1. Izquierda: Michael Wolgemut, Wilhelm Pleydenwurff, *Esferas celestes*, Xilografía iluminada. Tomado de: Hartmann Schedel *Crónica de Nüremberg* (1493). Derecha: *Cosmographia* de Pedro Apiano (1524).

Ambos grabados representan el cosmos aristotélico, como fuera adaptado de la obra de Ptolomeo por los padres de la Iglesia. Las esferas móviles se ubican más allá de las esferas del aire y el fuego terrestres, comenzando con la de la Luna y hasta la de Saturno. Luego vienen las estrellas y, envolviendo el todo, el empíreo inmóvil, «adonde habita Dios con los bienauenturados»[8]. Recordemos aquí que para los babilonios el universo era como una ostra esférica, donde la tierra flotaba sobre las aguas de las profundidades y estaba rodeada por las aguas superiores que recubrían la bóveda celeste. Las primeras serían visibles bajo la apariencia de fuentes naturales; las segundas, simplemente, como filtraciones del cielo en forma de lluvia. En la Edad Media, las aguas cósmicas superiores fueron retomadas por la exégesis bíblica que imponía que su existencia fuera real, aunque su verdadera naturaleza fuese discutida por los padres de la Iglesia, pues para algunos estas aguas se encontraban en estado fluido, pero otros las concebían más bien congeladas y duras como el cristal. De estas consideraciones surgió la esfera *Cristallinum* (el noveno cielo), ubicada entre el *Stellatum* o *Firmamentum* (las estrellas fijas, el octavo cielo), y el *Primum Mobile* o Primer Motor (el décimo cielo), lo que lleva el número de esferas a diez[9].

[8] Apiano, 1524.

Según el Pseudo-Dionisio Areopagita, los coros angélicos estaban a cargo del movimiento de los cielos. Este oscuro personaje neoplatónico del siglo V, probablemente un monje sirio, ejemplifica con su *Jerarquía Celeste*[10] la búsqueda de una explicación católica al complicado movimiento de las esferas de Aristóteles; para el Estagirita, el *Primum Mobile* impulsaba el movimiento de las esferas, que transmitían esta tracción a las capas inferiores. De más está decir que la visión religiosa podía hacerlo mejor. De acuerdo con el universo escalonado en esferas del Pseudo-Dionisio, los serafines hacían girar al Primer Motor y el *Stellatum* era movido por los querubines; los tronos, miembros del tercer coro de la jerarquía angélica, hacían lo propio con la esfera de Saturno. Los planetas Júpiter y Marte, y el Sol, correspondían, respectivamente, a las dominaciones, las virtudes y las potestades, mientras que los principados y los arcángeles se encargarían de las esferas de Venus y de Mercurio. La Luna, por último, estaba asignada a los ángeles inferiores. La aceptación de la jerarquía no fue inmediata pues luego de su traducción al latín por Juan Escoto Erígena, en el siglo IX, tuvieron que pasar tres siglos para que penetrara en la teología occidental[11]. Para el siglo XIV, Dante Alighieri (1265-1321) incorpora los preceptos del Pseudo-Dionisio en la *Divina Comedia*[12] (Fig. 2).

---

[9] Gangui, 2008.

[10] Hay tres jerarquías con tres coros cada una. La primera, integrada por serafines, querubines y tronos; la segunda por dominaciones, virtudes y potestades. La última jerarquía, formada por principados, arcángeles y ángeles.

[11] Le Goff, 2009, p. 134.

[12] Gangui, 2008.

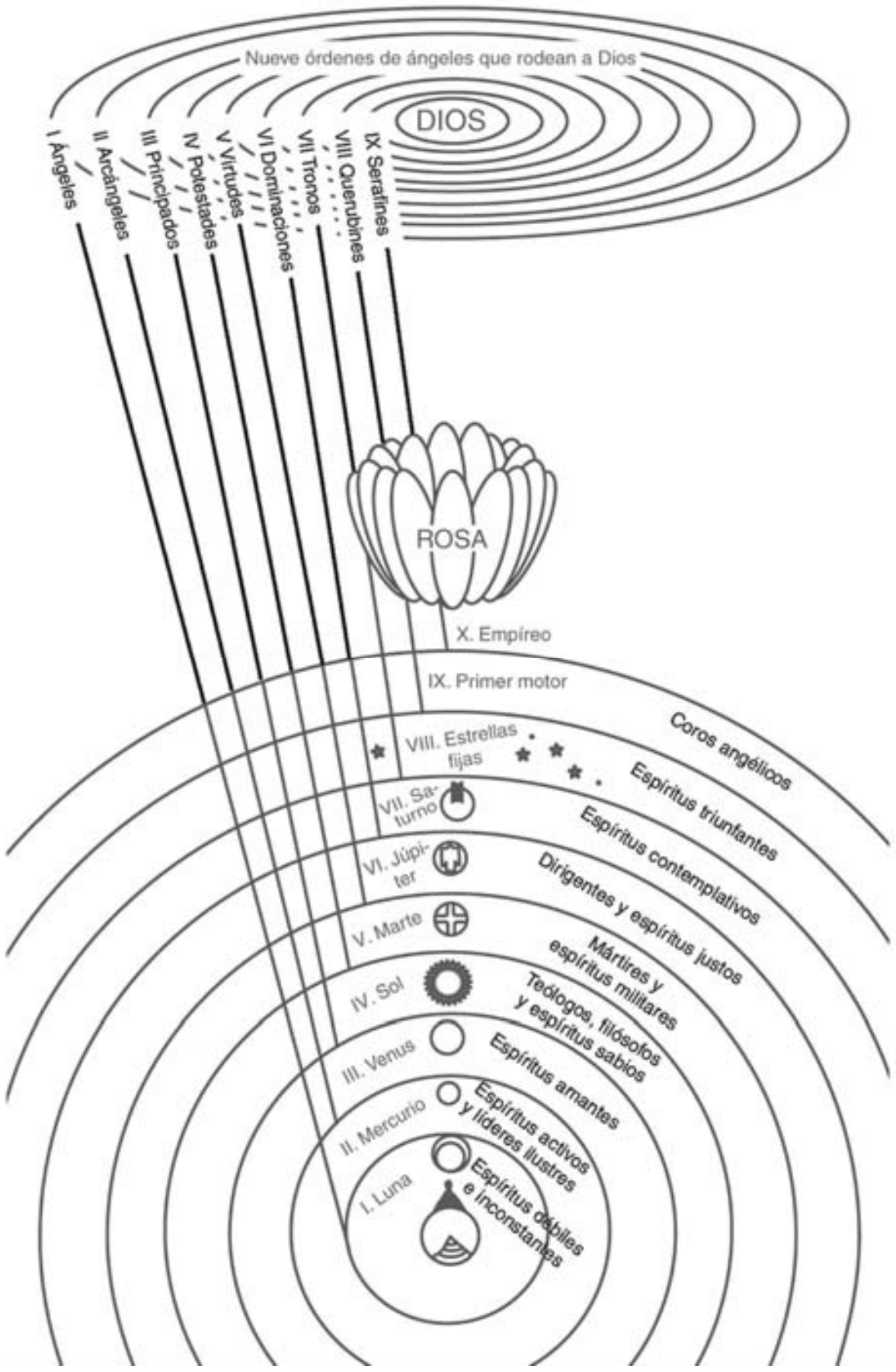

Figura 2. Esquema del cosmos de Dante.

Para Dante, los ángeles poseían un doble rol: contemplativo (de Dios) y activo, al tener a cargo el movimiento de los cielos (*Convivio*, tratado II, caps. 4-6). Las nueve esferas del Paraíso celestial, que lógicamente deberían ser un fiel reflejo de la contemplación (y no de la acción) contienen, sin embargo, la cruz en Marte, el águila del imperio en Júpiter y la escalera de la contemplación en el séptimo cielo de Saturno. Esta última era para Dante «...una escala suspendida / tan alto, que el final no se veía…» (Paraíso, canto XXI, 29-30), metáfora común de los escritos místicos que marca los pasos de un viaje espiritual hacia Dios. De otra parte Aby Warburg mostró que a las esferas de Dante se le sumaron también virtudes y saberes en el *cuattrocento* en una clave astrológica como muestra el *Salone* del *Palazzo della Ragione* de Padua[13].

[13] Aby Warburg Panel 23 del Atlas Mnemosyne. Tomado de Warnke, 2012, pp. 36-37.

## 2. Las ideas de ascenso y descenso del intelecto. Conocimiento científico y fe en Ramón Llull

De los modelos celestiales y su explicación por medio de las jerarquías del Pseudo Dionisio, se derivó un vínculo con la escala de Jacob. En la traducción latina comentada de la *Hierarchia Angelica* del siglo XVII se mencionó que dicha escala tenía dos columnas: «...la que sube simboliza la vida contemplativa y la que baja la vida activa»[14] (Fig. 3). Anteriormente lo que puede encontrarse en las versiones impresas de la obra del Areopagita son pirámides, que en varios grados confluyen en la divinidad. La escala de Jacob apareció en esta discusión sobre las esferas celestes y se impuso a la pirámide.

La exégesis de la dimensión simbólica de los textos religiosos se nutrió de la geometría y con su auxilio se establecieron complejos diagramas para compilar múltiples fuentes del conocimiento. La diagramática medieval recurrió al *stemmata* (del latín *stemma* y del griego στέμμα) que designa tanto una genealogía como una guirnalda o una corona, en cualquier caso un enlace. La visión analítica y memorística del *stemmata* implica jerarquías, órdenes, subordinaciones, que se graficaron en formas diversas siendo las más conocidas el árbol, la escalera y la rueda[15]. Por ello es adecuado el uso de la escalera para denotar el ascenso del espíritu hacia Dios, sobre la premisa de que todas las criaturas tienden a retornar a su creador.

El mallorquín Raimundo Lullio (o Ramón Llull, 1235-1315) vio los escalones faltantes que conducían desde la tierra hasta la jerarquía angélica del Pseudo Dionisio. Llull inicia su escala con los cuatro elementos, seguidos por la *jerarquía* del ser -desde los vegetales hasta el hombre- para terminar en el cielo con los ángeles y finalmente Dios[16]. Los nueve grados de esta escala son llamados por Llull *subiecta* (sujetos) a través de los cuales el entendimiento *asciende y desciende*[17]. Este es el fundamento de su *Liber de ascensu et descensu intellectus* (1305) cuya primera edición impresa (Valencia, 1512) contiene un grabado muy conocido con dos *stemmata* diferentes:[18] una escala arquitectónica con los nueve *subiecta* y un disco con dos ruedas concéntricas que tiene, en la exterior, la escala de doce *vocabulis* -inspirados en las categorías de la lógica aristotélica- y en la interior los cinco *gradus* para comprender la escala externa (Fig. 4).

[14] Le Goff, 2009, p. 134. En la traducción de 1634 de *Caelestis Hierarchia* se menciona la escala de Jacob en el capítulo XV: «Svsque: id est sursum deorsum, ut in Scala Iacob» [En ambos sentidos: esto es ascendente y descendente tal como la escala de Jacob]. Ver Areopagita (Corderio), 1634, p. 208. Sin embargo, en la edición impresa de 1502 no aparece esta alusión.

[15] Báez, 2005, pp. 34-40.

[16] Llull, [1304] 2002.

[17] Báez, 2005, pp. 80-83.

[18] Este grabado ha sido repetido en la obra de sus seguidores y en la edición de Mallorca de 1744.

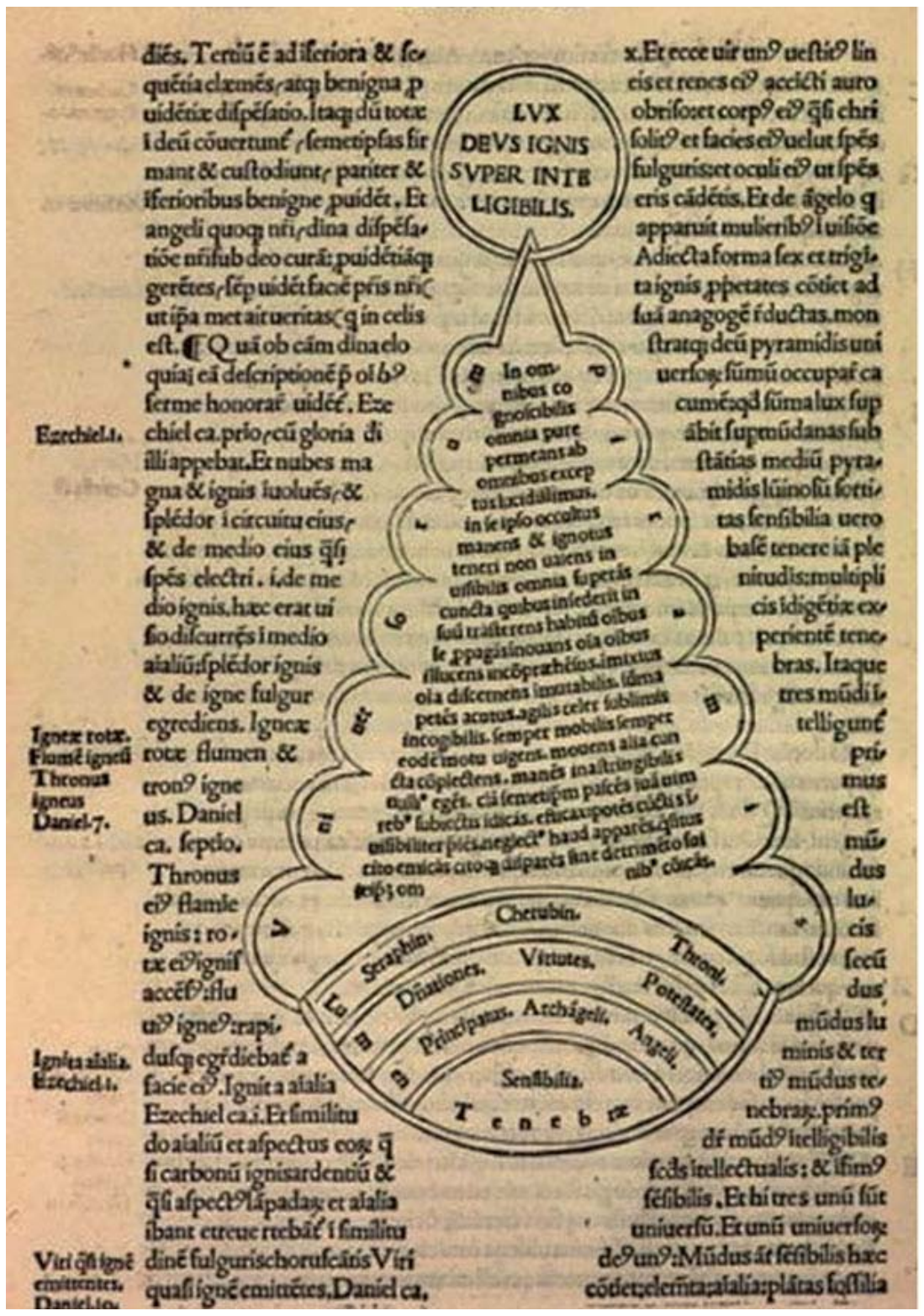

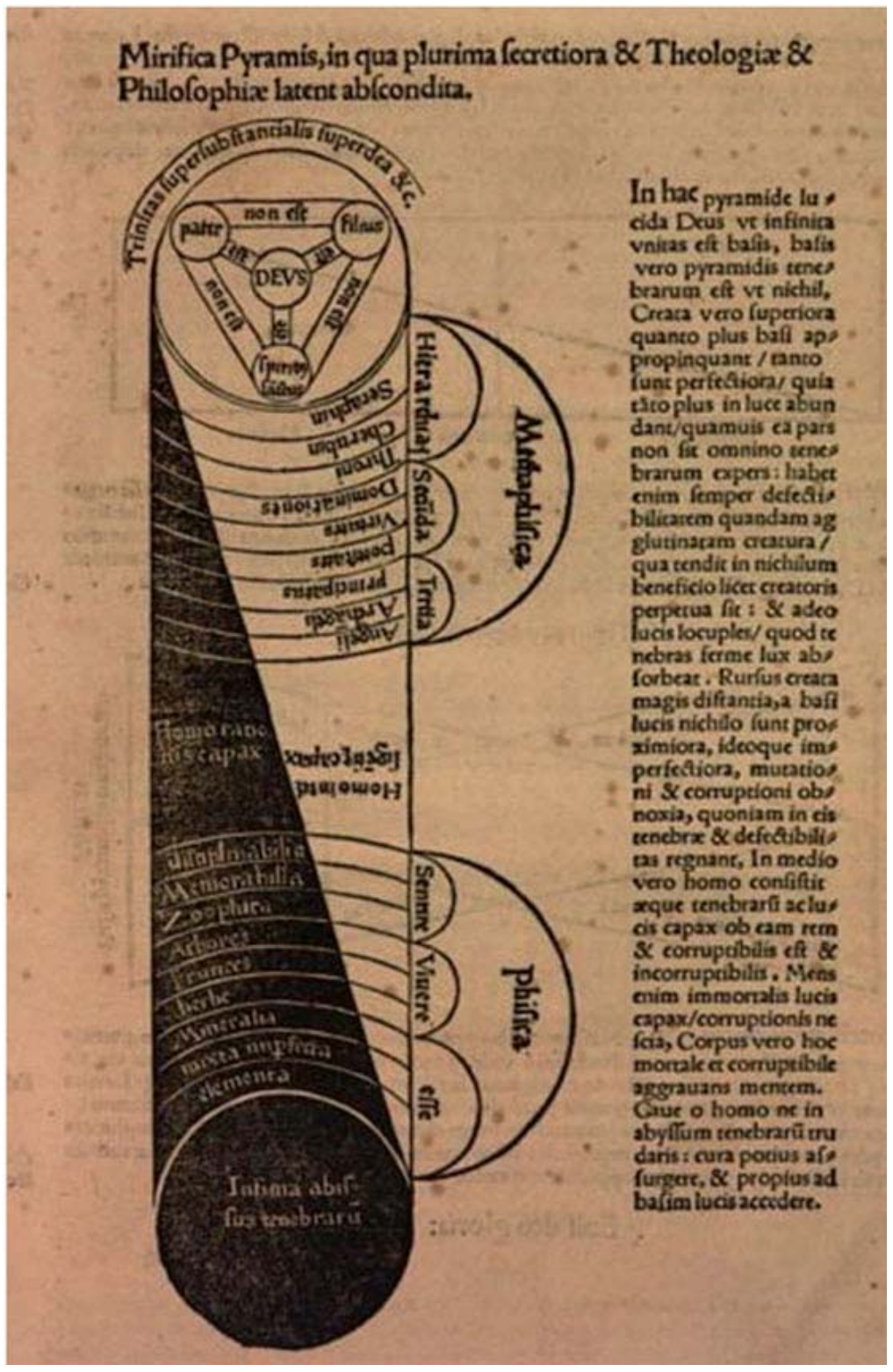

Figura 3. Pseudo Dionisio Areopagita. Izquierda: *Universorum Pyramis*. Tomado de: *Caelestis hierarchia. Ecclesiastica hierarchia. Divina nomina. Mystica theologia.* Joannes Tacuinus de Tridino, 1502, pp. XXIII v. Derecha: *Mirifica Pyramis*. Tomado de*: De mystica theologia liber I, graece…* Joannes Miller, 1519, pp. XXIII v. Biblioteca Nacional Austriaca.

Llull pretendía conciliar el pensamiento científico con la fe. Por ello se propuso mostrar que por todas las vías era posible demostrar la existencia divina, y que así se lograría convertir a judíos y musulmanes a la religión verdadera[19]. De ahí que buscara conciliar fuentes musulmanas y hebréas[20], el zodiaco con los *antiquis principiis Astronomiae* -los planetas, las esferas celestes-, los coros angélicos y las virtudes, para llegar a una *cábala cristiana*[21]. El resultado de este trabajo fue el influyente, aunque inédito, *Tractatus novus de Astronomia*, del que se conservan numerosas copias manuscritas[22]. Charles de Bouvelles (1479-1553), seguidor del pensamiento de Llull, retomó la escala del ser en su tratado *Physicorum Elementorum Libri decem* (1512), para componer la suya propia con 25 grados que van del *Mundus Sensibilis* al *Celestis* y culmina con Dios en el último grado que es el *Intellectualis*. Giordano Bruno (1548-1600), retoma las escalas de Llull en la edición que hizo Valerio Valeri de su obra[23], o las configura de manera diversa[24].

[19] Yates [1982] 1996, p. 19.
[20] Báez, 2005, pp. 68-75.
[21] Yates [1982] 1996, pp. 18, 31-54.
[22] Yates [1982] 1996, pp. 124-127.

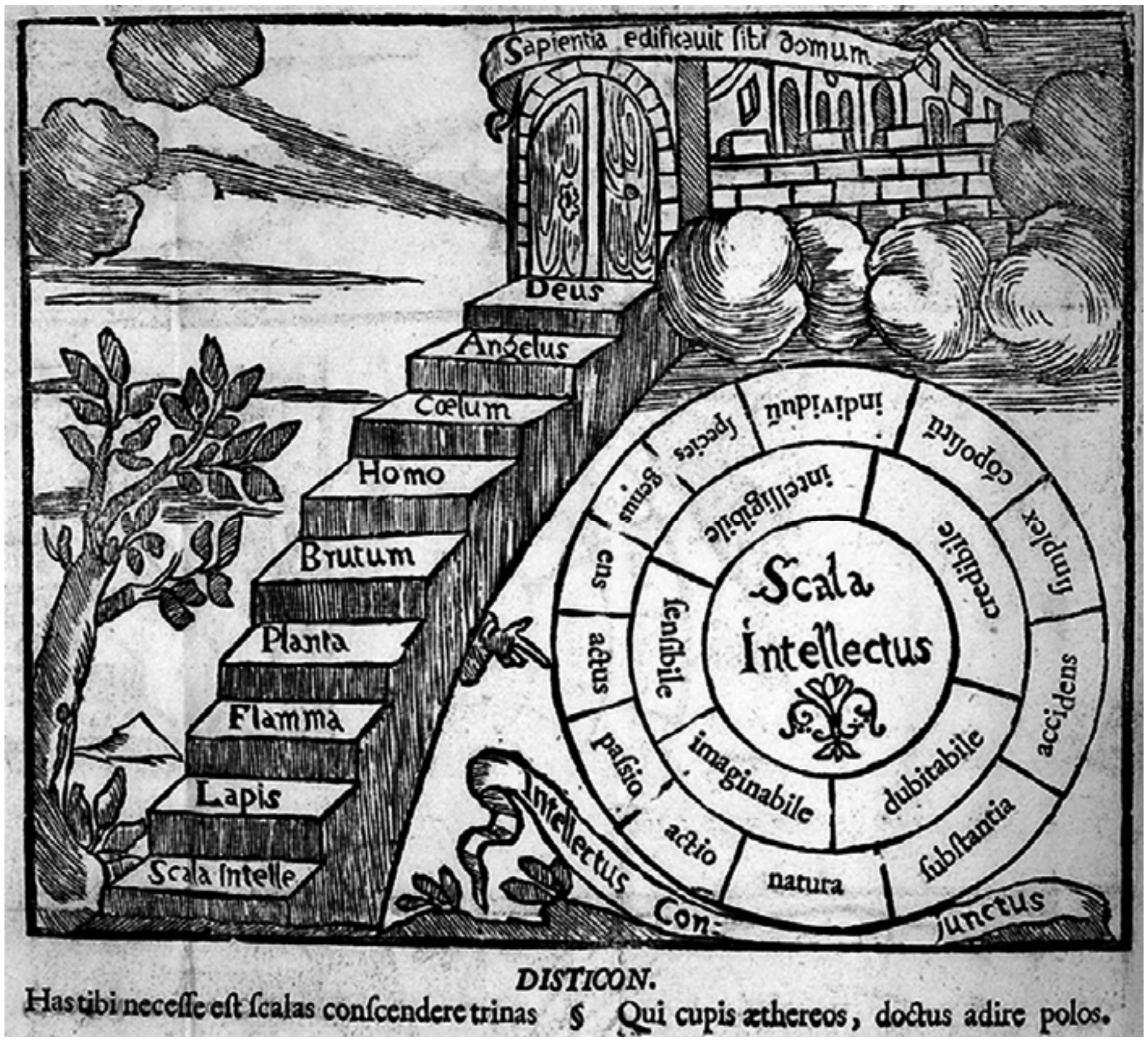

Figura 4. Raymundo Llull, *Liber de ascensu et descensu intellectus Palmae Majoricarum : ex typis Michaelis Cerdá & Arridi & Michaelis Amorós...*, 1744. Tomado de: Biblioteca Virtual del Patrimonio Bibliográfico http://bvpb.mcu.es

### 3. Apuntes para una iconografía de la escala

La visión de Jacob inspiró métodos graduales. En la *Regula* de Benedicto (c. 480-547) hay dos escalas, que pueden conducir al cielo por los doce grados de la *Humilitatis*[25], o bien al infierno por los grados de la *Supebiae*[26]. La *Scala Paradisi* de Juan Clímaco (siglo VII d.C.) estaba compuesta por treinta pasos de meditación hacia Dios. Este *stemmata* se extendió por toda Europa junto con la regla benedictina. Romualdo (c. 952-1027), benedictino fundador de los Camaldulenses, tuvo una visión en la que los monjes de su orden subían al cielo vestidos de blanco,[27] mientras que en Inglaterra el agustino Walter Hilton (1340-1396) creó la *Scala perfectionis* y, en la catedral de Bath –antiguo convento benedictino–, la escala está en la propia fachada, esculpida hacia 1520. Fuera del clero regular están los casos de la santa legendaria Catalina Virgen y martir[28] o la del manuscrito *Hortus deliciarum* (1167-1185) de Herrade de Landsberg, inspirada en la escalera de la virtud de Honorio de Autun (1080-1154)[29].

[23] Valeri, 1617. Esta obra suele ser atribuida erróneamente a Llull, cuando en realidad es una edición comentada por Bruno, Cornelio Agrippa y el propio Valeri.
[24] Bruno, 1582.
[25] Los grados son: 1. Timorem Dei; 2. Domini factis imitetur; 3. Obedientia se subdat Maiori; 4. Tacita conscientia patientiam; 5. Humilem confessionem Abbati; 6. Omni vilitate, vel extremitate contentus; 7. Humilians se..; 8. Nihil agat monachus.. nisi regula.. maiorum exempla; 9. taciturnitatem habens; 10. non sit.. promtus in Risu; 11. Loquitur monachus leniter & sine risu; 12. humulitate videntibus se Samper indicet.

Es así que la escala, como símbolo místico de ascenso espiritual, era bastante común, aunque en ninguno de los casos citados el cielo forma parte fundamental del esquema. Por ello es clave el aporte de Ramón Llull con su propuesta combinatoria de saberes, que usó a la escala como instrumento de la *ciencia revelada*. No hay que olvidar que Llull dispuso su *ars* a partir de una revelación del Espíritu Santo, por lo que para él no se trataba de un conocimiento especulativo sino de La Verdad. Por ello criticó la astrología antigua y la alquimia, que eran inferiores a su *medicina astrológica* (que inspiró los diagramas cósmicos de la *Medicina Catholica de* Robert Fludd).

[26] Cassiani, 1733.
[27] Hall, [1974] 1996: 324.
[28] Voragine, 1688: 618. Esta era homónima de las santas de Siena y Alejandría, mencionada en la *legenda 168*. Catalina formó con una cadena una escala de cuatro grados: 1. Inocentia Operis; 2. Mundicia Cordis; 3. Despectio Vanitatis y 4. Locutio Veritatis.
[29] Mâle, [1986] 2001: 137-138. Escalera compuesta por 15 peldaños: Patientia, Benignitas, Pietas, Simplicitas, Humilitas, Contemptus Mundi, Paupertas Voluntaria, Pax, Bonitas, Spirituale Gaudium, Sufferentia, Fides, Spes, Longanimitas, Perseverantia.

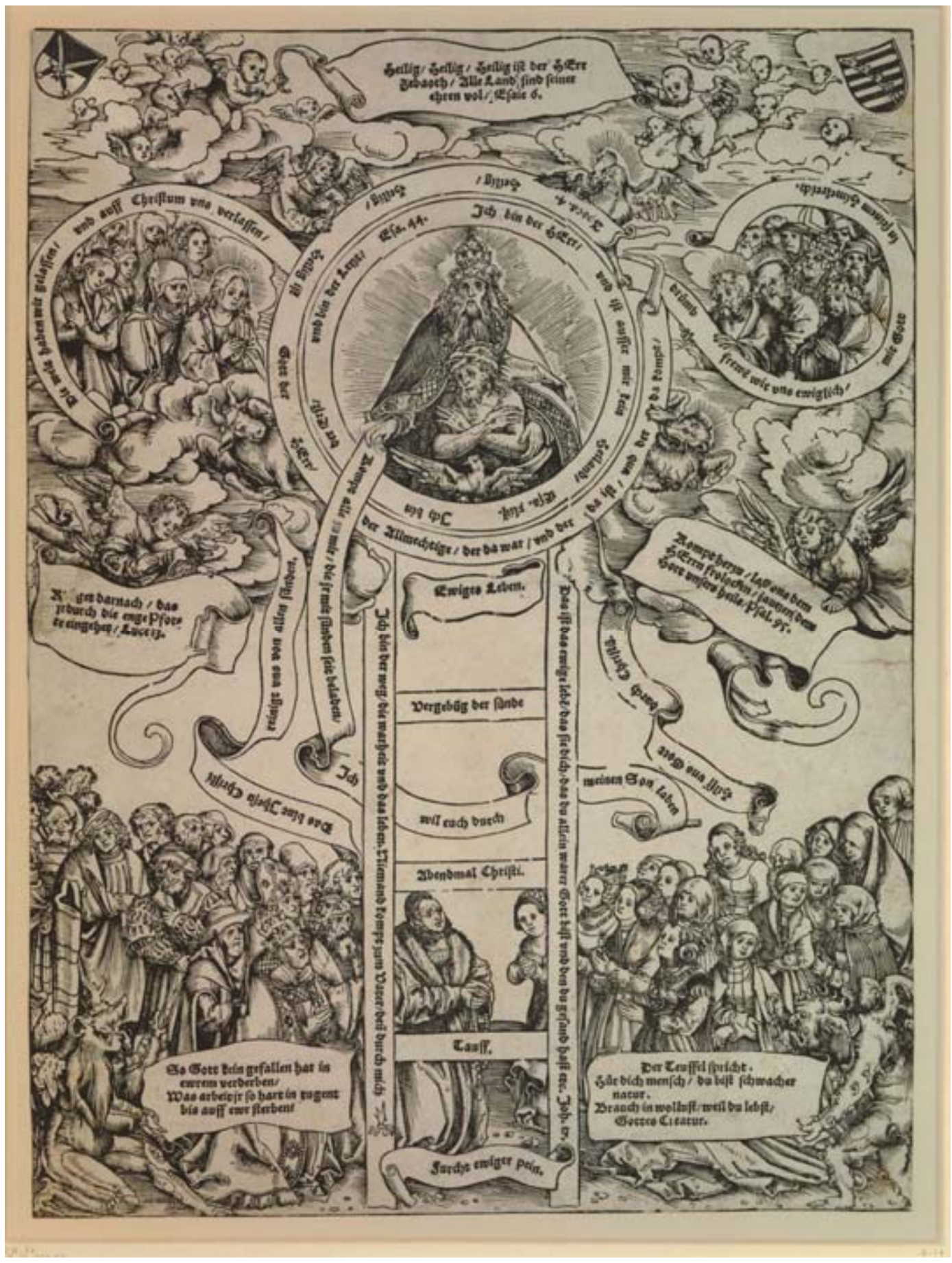

Figura 5. Lucas Cranach, el viejo (c. 1472-1553) / Symphorian Reinhart, imp. (activo c. 1509) *Escalera celeste de San Buenaventura*, c. 1508-1511. Xilografía © Trustees of the British Museum.

La llegada a América de las escalas se relaciona con las órdenes religiosas. El libro de exempla *Scala Coeli* (1485) del dominico Johannes Gobius (c. 1300-1350), fue usado en Nueva España para escribir los *tlamahuizolli* contenidos en el *Códice 7* en náhuatl, conservado en la John Carter Brown Library[30]. De otra parte los autores místicos del siglo XVII hicieron escalas como la de María de Ágreda (1602-1665), monja Concepcionista, y la del agustino descalzo Agustín de San Ildefonso (muerto en 1662), en este caso grabada por Claudio Coello[31]. Para el Alto Perú es conocido el uso de los coros angélicos en el *Coloquio de los once cielos* del monasterio de Santa Teresa de Potosí[32]. Si bien en esta obra no hay escalas, sí hay relación con algunas composiciones del *Cancionero Mariano de Charcas* donde se alude a la Virgen como escala celestial: «...cantad que la escala es María», «...Salve, Aurora, et Scala Salve» o directamente *escala de Jacob*, como aparece en un verso del *Nombre de María*.[33] Pero para el tema que nos ocupa, más pertinente aún es la *Copla a la Concepción* donde la Virgen es alabada con metáforas celestes, en las que interviene también la escala:

> Reina, escala, cristal, día, / Reina de uno y de otro orbe, escala sin leve riesgo,
> cristal sin nunca empañarse, / día sin noche, naciendo;
> Reina, escala, cristal, día / eres de este firmamento[34]

[30] Rubin, 2009: 34-35.
[31] San Ildefonso, 1683: s. p.
[32] Eichmann, 2003: 95-132.

Estas composiciones aluden al ciclo iconográfico de la *Tota Pulchra*, que interpreta en el *Cantar de los Cantares* y en el *Génesis* menciones a la Virgen como puerta del cielo, escala de Jacob, estrella matutina, Luna y Sol.[35]

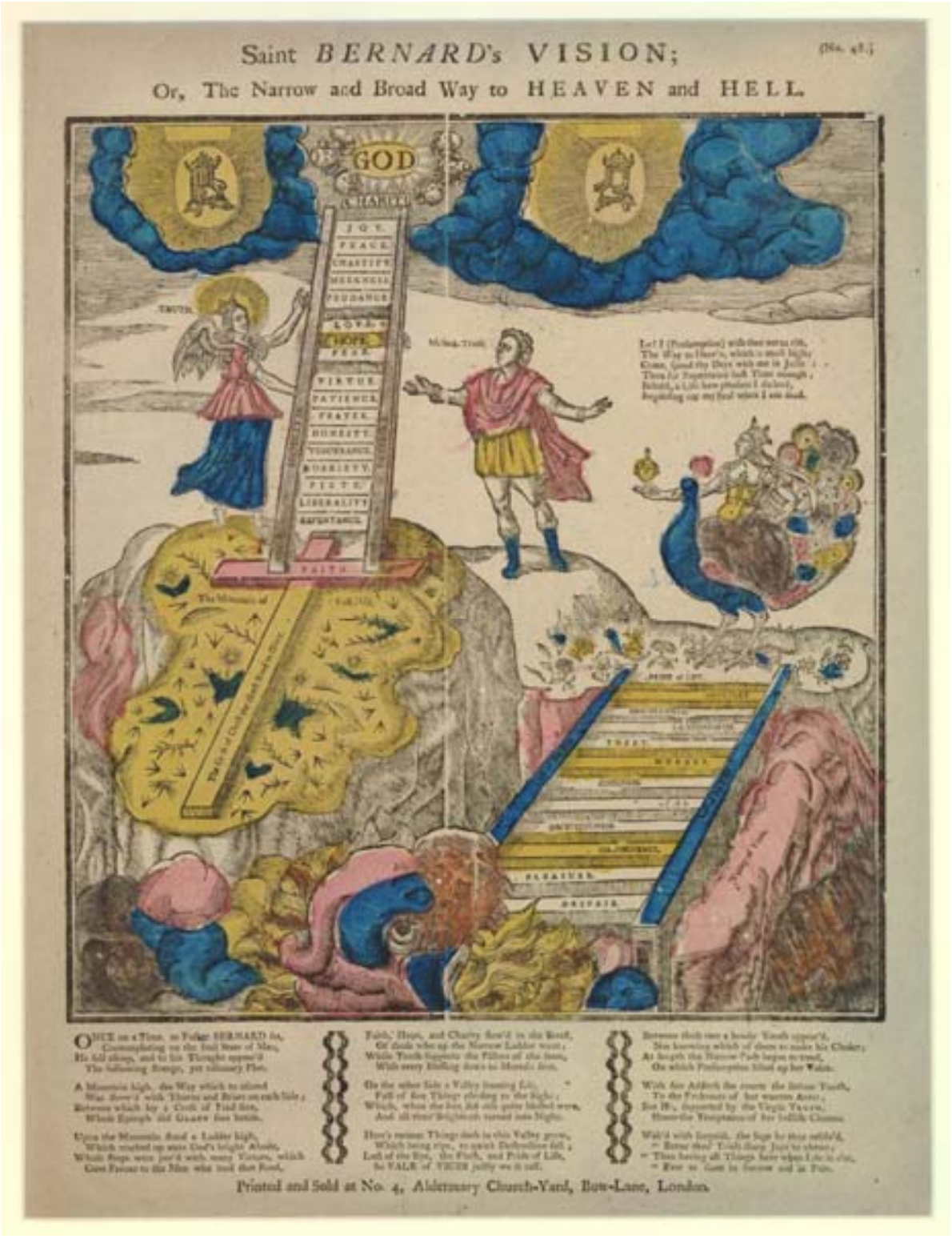

Figura 6. Anónimo *Saint Bernard's Vision*, c. 1750. © Trustees of the British Museum. Versión de Bonifacio Natale (1538-1592) grabó, Claude Duchet (1570-1585) editor. *Scala coeli et inferni ex divo Bernardo* c. 1581, conservada en la Biblioteca Nacional de España.

[33] Eichmann, 2009, pp. 363, 581, 705-706.
[34] Eichmann, 2009, pp. 272-273.
[35] Schenone, 2008, pp. 28-33.

## 4. Un *stemmata* celeste

En el ámbito Altoperuano las escalas se conocían en relación con la iconografía mariana, aspecto que en la obra que analizamos no se destaca, como sí lo hacen en cambio las escalas celestes y los cielos astronómicos y metafísicos. La complejidad de los detalles constitutivos del cuadro muestra claramente que su objetivo principal no fue la conversión de fieles, pues no trata un tema catequético. Además, el nombre de la obra no se ha aclarado aún, aunque se ha identificado como perteneciente a la escuela cuzqueña del siglo XVIII bajo dos extensos títulos dados por sus investigadores: *Alegoría de la subida al Trono Divino con diagrama cosmológico de los cuatro elementos, planetas, cielos y jerarquías angélicas*[36] y *Alegoría del firmamento con la representación de los siete cielos de los planetas, más el círculo de estrellas fijas donde está el zodiaco, más el «cristalino», el «primer móvil» y el «empíreo»*[37] (Fig. 7).

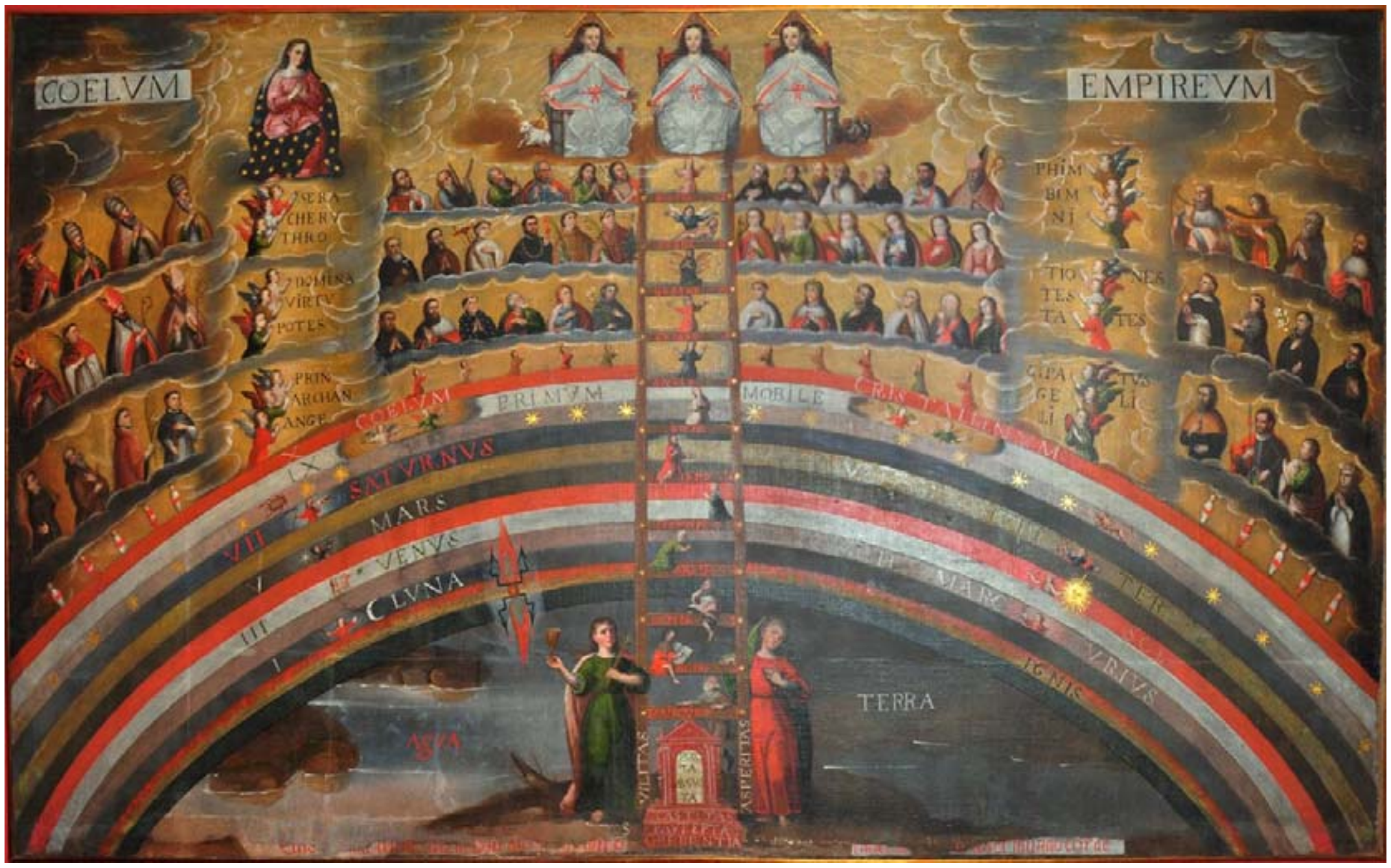

Figura 7. Anónimo. *Alegoría del firmamento*, Cuzco, siglo XVIII.

Sin embargo, la existencia de -al menos- tres obras de características muy similares a la cuzqueña indica que esta iconografía posiblemente se propagó a través de la estampa. Las obras en cuestión son: una pintura en el monasterio franciscano de Kalwarya Zebrzydowska,

[36] Mujica Pinilla, 2002.
[37] Gisbert, 2004.

Polonia[38] y dos españolas, una perteneciente a la parroquia de Santa María de Dueñas, en Palencia,[39] relacionada por primera vez con la obra cuzqueña en 2010,[40] y otra en la Iglesia del Carmen de Antequera[41]. La semejanza entre las cuatro obras es notable, a pesar de que la obra cuzqueña es apaisada a diferencia de las europeas, pues todas tienen los mismos elementos principales, con ligeras diferencias formales e iconográficas (Fig. 8).

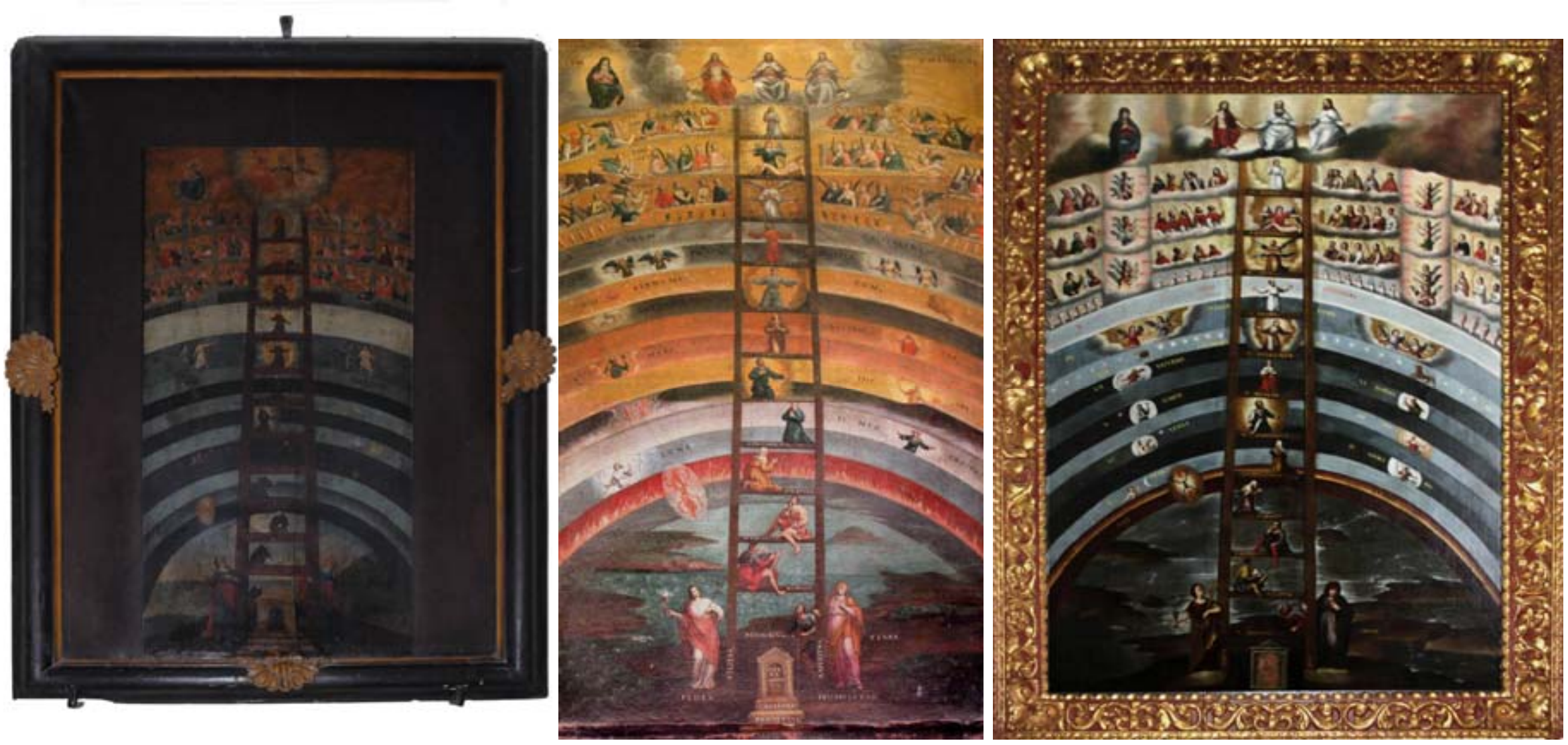

Figura 8. Versiones europeas de la *Escala celeste*. A. Monasterio franciscano de Kalwaria Zebrzydowska, Polonia B. Iglesia de Santa María de Dueñas (Palencia, España) C. Iglesia del Carmen (Antequera, España). Fotografías cortesía de Ewa Kubiak y de Arturo Caballero.

Las obras están divididas en los mundos sublunar, astral y empíreo, con los cuatro elementos dispuestos sobre una porción de tierra rodeada de agua, TERRA y AQUA, sobre la que están los elementos restantes: AER y, junto a una franja de fuego, IGNIS. Se complementa sobre la izquierda con la representación del rayo, que es el atributo tanto del fuego como de Júpiter.[42] Las esferas celestes siguen a continuación, en arcos concéntricos, con el astro y el dios que las identifica y sus atributos, todas señaladas con números romanos en orden ascendente. Después siguen los tres arcos correspondientes al Firmamento –donde se ubican las estrellas fijas y el zodiaco–, el Primer Motor y, por último, el Cielo Cristalino (ver Fig. 9). Llamativamente, en el primero aparecen solamente los signos de Cáncer, a la izquierda, y de Leo a la derecha, lo cual tiene un sentido simbólico. Según Llull, Cáncer representa «…nocturnidad, feminidad, y su planeta es la luna, y es de la complexión del agua», de otra parte, Leo es «motividad, masculinidad, diurnidad y su planeta es el sol; y […] es de complexión de fuego».[43] Adicionalmente, Cáncer y Leo apuntan al cielo del sur o *Meridies*, que en la representación simbólica de la Jerusalén celeste coincide con la obra estudiada. Según Charles de Bouvelles (1479-1553) en el *Meridies* o sur, aparecen las jerarquías angélicas más tres grados terrestres: Pontífices, Sacerdotes y Ministros, a los que se corresponden los siete planetas de la época, más el Firmamento y los cuatro elementos[44].

[38] Cichon y Kubiak, 2012. Agradecemos la generosidad de Ewa Kubiak que nos permitió enriquecer el presente trabajo.
[39] Caballero, 1988.
[40] Vanegas Carrasco, 2010. En este trabajo se lleva a cabo un primer acercamiento a las escalas europeas.
[41] Cichon y Kubiak, 2012.
[42] Hall, 1996, p. 319.

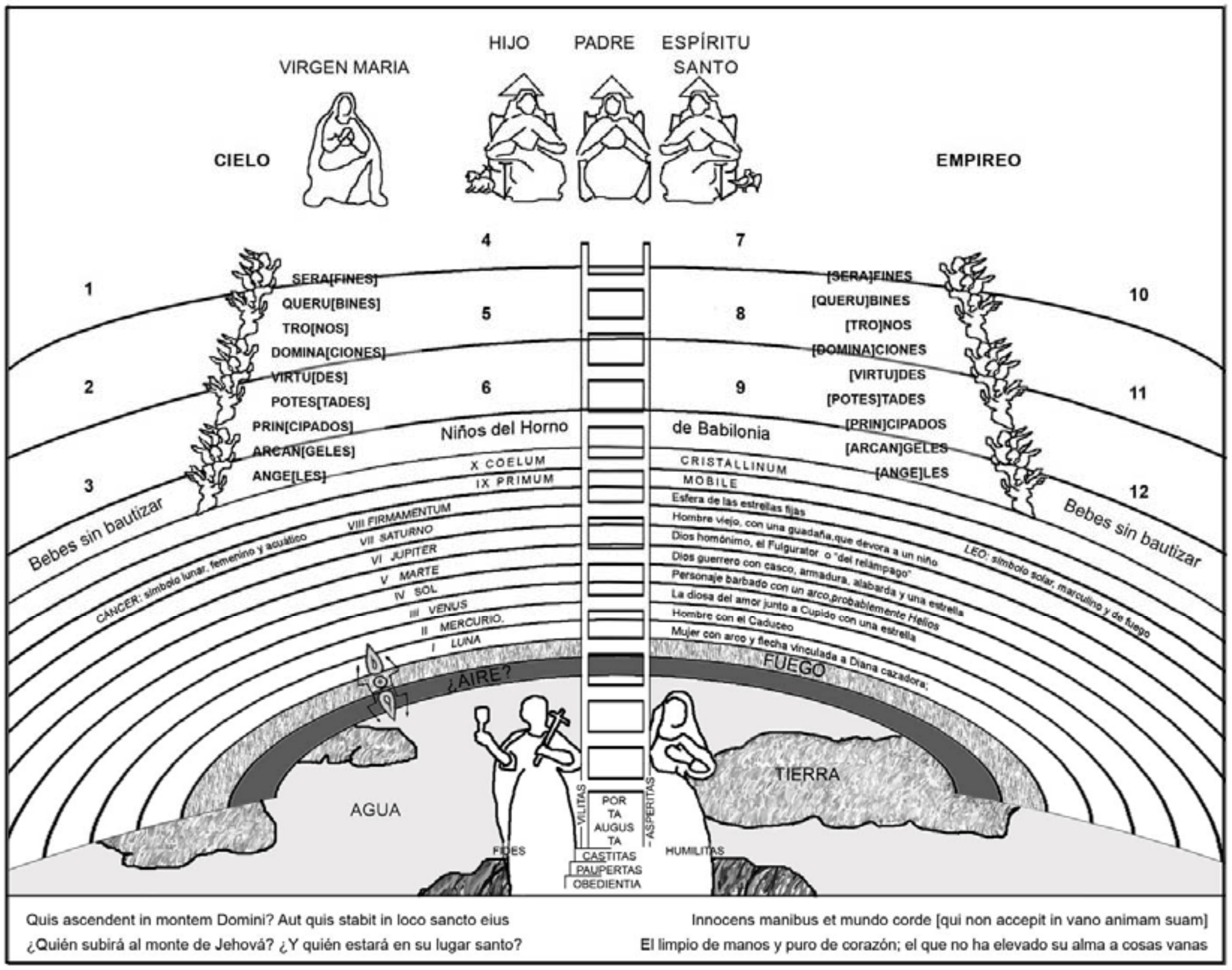

**Habitaculum Dei et omnium electorum**

1. Un Cardenal y tres Papas sin atributos (¿Los cuatro Doctores de la Iglesia?)
2. S. I. (¿San Alberto Magno?), Mártir S. I., S. I., S. I.
3. ¿San Juan de Dios?., Simón Stock, Mártir S. I., S. I.
4. S. I. (¿Evangelista?), San Andrés, San Pablo, San Pedro, San José, San Juan Bautista
5. San Antonio el Grande, S. I., ¿San Agustin?, San Bernardo de Claravall, San Esteban Protomártir, San Lorenzo
6. S. I., ¿San Luis Rey de Francia?, San Nicolás de Tolentino, S. I., S. I. (¿San Joaquin y Santa Ana?), San Antonio de Padua
7. S. I. (¿Santo Tomás?), Santo Domingo, San Francisco, San Pedro Nolasco, ¿San Ignacio de Loyola?, San Francisco Javier, S. I. (¿San Buenaventura?)
8. Santa Úrsula, Santa Bárbara, Santa Catalina de Alejandría, Santa Inés, S. I., S. I.
9. S. I., S. I., S. I., Santa Teresa de Ávila, Santa Clara de Asís, S. I.
10. David, ¿Noé?, S. I., S. I.,
11. San Jacinto, Santo franciscano, ¿San Juan Damasceno?, ¿San Benito de Palermo?
12. Peregrino S. I.(¿Santiago el Mayor?), ¿Longinos? S. I. ¿Santa Rosa de Lima?

**S. I.: Sin Identificar**

Figura 9. Esquema detallado de *Alegoría del firmamento*, con sus diferentes partes, junto a la posible identificación de los personajes sagrados y alegóricos de la escala celeste, los cielos astronómicos y las jerarquías celestes. Dibujo J.R.Rey.

[43] Llull, [1304] 2002, p. 96.
[44] Bouvelles, 1510.

En el *Primum Mobile* hay dos parejas de ángeles sosteniendo relojes de arena, que podrían indicar que a partir de ese punto se ingresa donde se suspende el tiempo físico y empieza la eternidad[45]; a este sigue el cielo cristalino (o *Coelum Cristalinum*). Ahora bien, en las obras estudiadas se sitúa en la novena esfera el Primer Motor y en la décima el Cristalino, orden que se invierte en la representación de la *Cosmographia* de Apiano, mientras que en la representación de Dante no se evidencia el Cristalino y se ubica en el noveno cielo al Primer Motor. Quizá por este motivo Teresa Gisbert sigue a Dante e identifica el contenido del cuadro con el *Coloquio de los once cielos*[46] y con *El tratado del cielo* del jesuita José de Aguilar (Sevilla, 1701)[47]. Sin embargo, Dante menciona solo diez cielos y el Cristalino *es* importante en la obra que nos ocupa, pues encima de él hay una hilera de niños arrodillados, seguidos por un grupo de bebés fajados. Este espacio puede representar el limbo[48] donde están «los niños del Horno de Babilonia y los que murieron sin bautizo».[49]

La tercera sección de las obras corresponde al Empíreo, *Habitaculum Dei et omnium electorum*, y es la parte más llamativa del cuadro cuzqueño. En lo que respecta a los personajes de las jerarquías celestes, a pesar de la dificultad de identificarlos a todos, en el lienzo cuzqueño son más reconocibles que en las otras obras[50], con las que existen notorias diferencias tanto en orden como en cantidad (ver Fig. 9). En el *Empirevm* está la Trinidad, figurada como tres personas, con la Virgen María entronizada (incluso coronada en la pintura polaca). Este detalle, más que un improbable error iconográfico[51], muestra una postura frente a los debates marianos y el lugar que la madre de Cristo había de ocupar junto a su hijo.

Nos concentraremos ahora en la parte central de las obras, donde se reúne la iconografía cosmológica con la de las escalas, para otorgarle un sentido particular a las obras estudiadas. Para un esquema de los escalones, con sus mínimas diferencias, ofrecemos un cuadro adicional (ver Fig. 10). Lo que más interesa en este punto es aquello que nos indica, como en el caso de la Virgen, la complejidad del entramado cultural de una época, para lo cual fue indispensable comparar las cuatro obras para comprender el programa iconográfico, ventaja con la que no contaron autores anteriores[52]. En la parte inferior central hay tres escalones OBEDIENTIA, PAVPERTAS, CASTITAS, bajo una puerta identificada como «PORTA AVGVSTA» («ANGVSTA», en el caso de Antequera) por donde se ingresa a la vida espiritual. A los lados de la puerta están las alegorías «FIDES» y «HVMILITAS», identificadas con inscripciones apenas legibles en la obra cuzqueña. Los parales de la escalera tienen dos inscripciones: «VILITAS», a la izquierda, y «ASPERITAS», al otro lado, que sostienen doce peldaños identificados con una alegoría y un concepto, que en todas las obras coincide parcialmente. La claridad de las inscripciones es menor en la pintura polaca, mientras que en la cuzqueña se presentan errores de escritura, que podrían adjudicarse a la copia de la estampa, pues los grados coinciden con los de la obra palentina.[53]

[45] Caballero, 1988.
[46] Eichmann, 2003.
[47] Gisbert, 2004; Gisbert y Mesa Gisbert, en Siracusano, 2010, pp. 15-30.
[48] Caballero, 1988, p. 264.
[49] Gisbert y Mesa Gisbert, 2010, p. 26.
[50] Caballero, 1988, p. 266.
[51] Caballero, 1988, p. 267.

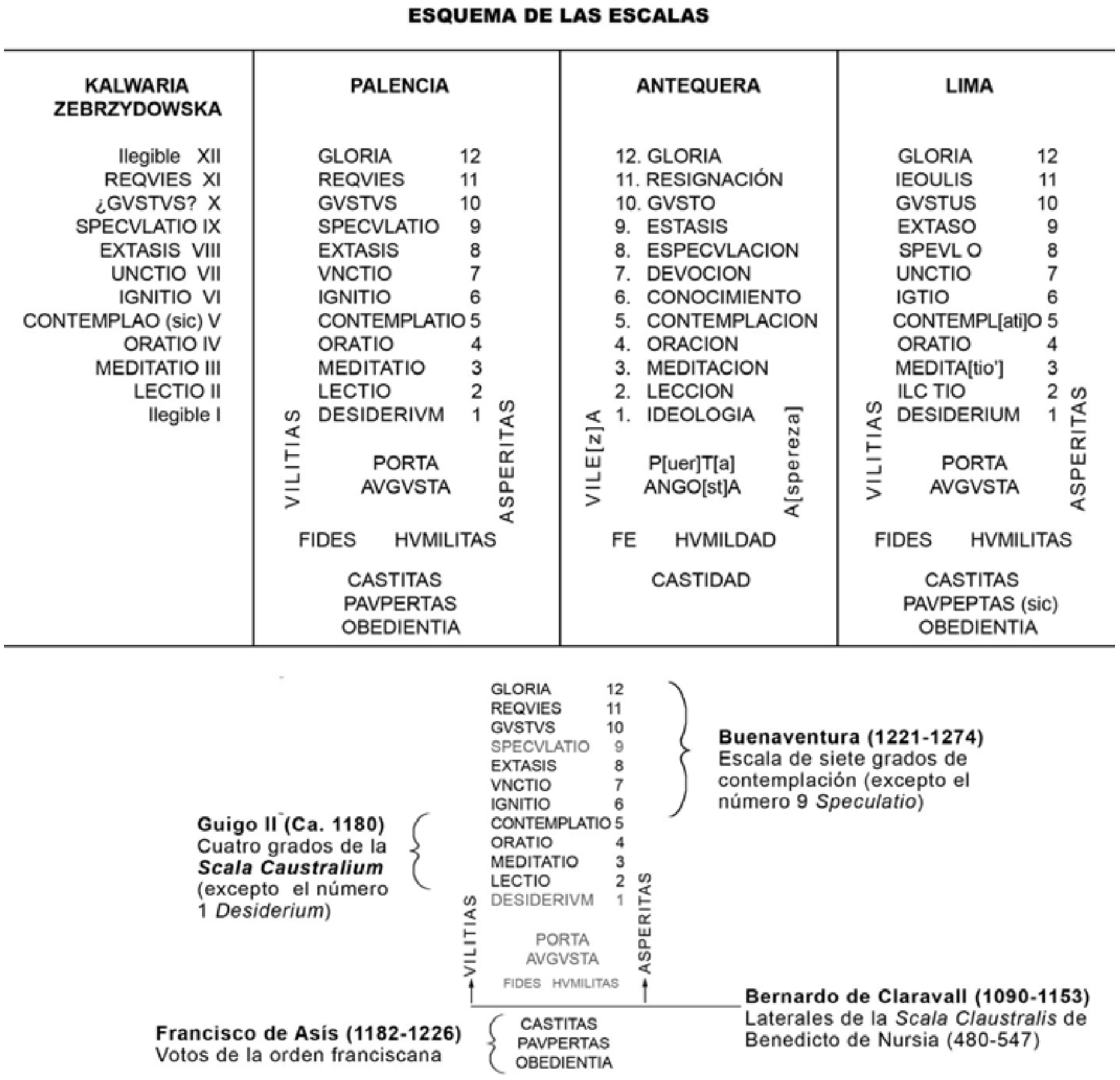

Figura 10. Esquema de los conceptos legibles en los peldaños de las cuatro obras conocidas hasta el momento.

[52] Mujica Pinilla, 2002; Gisbert y Mesa Gisbert, 2010.

[53] Los errores a los que se alude son, en primer lugar, de contracción, en: 03. MEDITA (meditatio), 05. CONTEMPLO (Contemplatio), 06. IGTIO (Ignitio) y 08. SPEVL O (Especulatio). De otra parte están los errores de escritura como en 09. EXTASO (Extasis), 02. ILC TIO (Lectio) y -quizá el más complicado- 11. IEOULIS que debería decir REQUIES.

## 5. Una escalera entre tierra y cielo, que une Europa y América

Terminado este análisis sucinto de la obra, estamos en condiciones de proponer una interpretación alternativa a las presentadas en trabajos previos. Como se mencionó, nos encontramos frente a un *stemmata* sobre la virtud en la vida monástica. Según Bernardo de Claravall (1090-1153) *Vilitas* y *Asperitas* son las letras de una *Scala Claustralis* que tiene como grados las virtudes y las gracias propuestas por Benedicto en su regla.[54] *Vilitas* y *Asperitas* aluden a lo vil y áspero de la vida monacal como escribió el cartujo Guigo II, en una carta en la década de 1180, donde menciona una *Scala claustralium* compuesta por cuatro grados: lectio, meditatio, oratio y contemplatio[55]. Poco después Buenaventura (1221-1274) en sus *Commentarius in Evangelium S. Lucae*, menciona una escalera de siete grados de la contemplación: fuego, unción, éxtasis, contemplación, gusto, descanso, gloria.[56] Es decir, en total once grados que coinciden de manera exacta con los peldaños representados en nuestra escala mística: del 2 al 5 de Guigo y del 6 al 12 de Buenaventura, más los laterales de Bernardo de Claravall, con excepción del número 9, «Expeculatio», que en los *Commentarius* de Buenaventura es «Contemplatio». Si a esto le sumamos que los tres peldaños bajo la escala forman parte de los votos de la orden franciscana, se hace evidente una preocupación por la enseñanza metódica de la virtud, iniciada con la regla de Benedicto en el siglo VI, continuada en el siglo XII por Guigo y Bernardo de Claravall, y culminada la centuria siguiente por Buenaventura. Estas ideas circularon en el mundo monástico a través de los siglos pues la escala de Guigo fue retomada por Girolamo Savonarola (1452-1498) como grados de oración.[57] Probablemente los escalones representen las tres vías místicas de Juan de la Cruz: purgativa (grados 1 al 4), iluminativa (del 5 al 8) y unitiva (del 9 al 12).[58] Arturo Caballero identificó la escala mística de Palencia con los escritos del sacerdote español Miguel de Molinos (1628-1696), autor de *Manuductio Spiritualis* [Guía espiritual] (1675), condenado por la Inquisición española.[59] La identificación de las obras estudiadas con una fuente herética es llamativa pero carece de sustento, pues Molinos propuso múltiples escalas, entre ellas la de *Contemplationis*, similar a la de Buenaventura.[60]

[54] Haec ergo sint latera scalae, vilitas et asperitas: quibus deinceps internae virtutis et gratiae gradus firmiter inserantur. Claravall, [s. XI] 1879, pp. 460-462.
[55] Hexter y Townsend, 2012, p. 481.
[56] Canty, 2011, p. 180.
[57] Savonarola, 1504.
[58] Caballero, 1988, p. 273.
[59] Caballero, 1988, p. 273. Molinos, autor del «quietismo», fue criticado por los jesuitas Gottardo Bell'Uomo y Paolo Segneri, y por el cardenal D´Estrees, embajador en Roma de Luis XIV. Fue denunciado ante la Inquisición en 1678, por lo que redactó su *Defensa de la contemplación* (1679-1680). Procesado en 1685 por inmoral y heterodoxo, fue condenado a reclusión perpetua, abjuró de sus errores en 1687, falleciendo en 1696.

La existencia de cuatro obras sobre el mismo tema claramente no es casual, y abre el interrogante sobre una procedencia común, como estampa suelta o frontispicio de un libro, con lo cual podría haber más representaciones de la escala mística. Sería interesante dilucidar si las diferencias entre las obras son deliberadas o casuales, cuestión que nos remite al proceso de la copia y las necesarias reinterpretaciones de la fuente, ya sea por razones intelectuales, técnicas o de producción. Este último punto adquiere relevancia a la luz de un último descubrimiento que surgió en nuestro abordaje. En la zona inferior de la escala cuzqueña hay una inscripción que la diferencia de las demás obras. Se trata de un texto, parcialmente borrado, en el que pudimos reconocer una cita bíblica dividida en dos frases dispuestas a ambos lados de la base de la escalera: a la izquierda «Quis ascendent in montem Domini? Aut quis stabit in loco sancto eius», y a la derecha «Innocens manibus et mundo corde». La cita está cortada, pues continúa con la frase «...qui non accepit in vano animam suam, nec Iurabit in dolo proximo suo».[61] Se trata del salmo 24, cuya presencia junto al esquema presentado evidencia que estas imágenes pudieron ser creadas para ser observadas dentro de círculos intelectuales en Europa, pero también en América.

Nuestro trabajo muestra la correspondencia entre las escalas y una idea del cosmos en la pintura cuzqueña, dando así testimonio de un debate que fue iniciado en el medioevo. Pero su actualización en el ámbito conventual europeo y americano, muestra cómo el misticismo de los siglos XVI y XVII reunió el esquema cosmológico y el *stemmata* de las escalas; una apropiación y continuidad en un debate antiguo, al margen de las necesidades de evangelización del Nuevo Mundo.

## BIBLIOGRAFÍA

Apiano, P. [1524] *La Cosmographia de Pedro Apiano*. Anvers, 1575, ed. Biblioteca de la Universidad de Sevilla, s. f.

---

[60] Molinos, 1687: 356. Los grados son: ignis, unio (*sic*), elevatio, illuminatio, gustus y requies.

[61] ¿Quién subirá al monte de Jehová? ¿Y quién estará en su lugar santo? El limpio de manos y puro de corazón; El que no ha elevado su alma a cosas vanas, Ni jurado con engaño. En la *Biblia Vulgata* el salmo tiene el número 23.

Areopagita, Pseudo Dionisio, Baltasar C. *Opera S. Dionysii Areopagitae cum scholiis S. Maximi et paraphrasi Pachymerae a Balthasare Corderio latine interpretata* Amberes, Officina Plantiniana, 1634.

__ . *Caelestis hierarchia. Ecclesiastica hierarchia. Divina nomina. Mystica theologia.* Venecia, Joannes Tacuinus de Tridino, 1502.

__. *Obras Completas: Los nombres de Dios. Jerarquía celeste. Jerarquía eclesiástica. Teología mística. Cartas varias*, Madrid: Biblioteca de Autores Cristianos, 2002.

Báez, L. *Mnemosine novohispánica,* México, UNAM, 2005.

Bouvelles, C. *Physicorum Elementorum Libri decem.* Ioannis Parui & Iodocii Badi, 1512.

Bruno, G., *De umbris idearum implicantibus artem quaerendi, inveniendi, indicandi*, *ordinandi & applicandi*. Paris, Aegidium Gorbinum, 1582.

Caballero, A. «Un itinerario místico por el cosmos», Palencia, Publicaciones de la Institución Tello Téllez de Meneses, 1988, núm. 58, pp. 251-280.

Canty, A. *Light and Glory: The Transfiguration of Christ in Early Franciscan and Dominican Theology* Michigan: Catholic University of América Press, 2011.

Cassiani, J. *Opera omnia, cum commentariis D. Alardi Gazaei.* Wetstenios et Smith, 1733.

Cichon, K. y Kubiak, E. «Entre la tierra y el cielo»,*Quaderni di Thule*, *Atti del XXXIV Convegno Internazionale di Americanistica*, vol.12, Circolo Amerindiano «Onlus», Perugia 2012.

Claravall, B. de, *Opera omnia: sex tomis in quintuplici volumine comprehensa posthorstium denuo recognita...* París: Garnier fratres, 1879.

Demaray, J. G. *Dante and the Book of the Cosmos.* Filadelfia, American Philosophical Society, 1987.

Duby, G. *Arte y sociedad en la edad media*, Madrid: Taurus, 2011.

Eichmann, A. «El coloquio de los once cielos. Una obra de teatro breve del Monasterio de Santa Teresa (Potosí)», *Historia y Cultura*, núm. 28-29, La Paz, Sociedad Boliviana de Historia, 2003.

__, *Cancionero mariano de Charcas*. Madrid / Frankfurt: Iberoamericana/Vervuert, 2009.

Gangui, A. *Poética Astronómica: El cosmos de Dante Alighieri.* Buenos Aires, Fondo de Cultura Económica, 2008.

Gisbert, T. *El paraíso de los pájaros parlantes. La imagen del otro en la cultura andina*. La Paz: Plural ediciones, 2001.

__. «El cielo y el infierno en el mundo virreinal surandino», en *Memoria del II Encuentro Internacional. Barroco y fuentes de la diversidad cultural*. La Paz: Viceministerio de Cultura, Unión Latina, Unesco, 2004.

Hall, J., [1974] *Diccionario de temas y símbolos artísticos*. Madrid, Alianza, 1996.

Hexter, R. y Townsend, D. *The Oxford Handbook of Medieval Latin Literature* New York: Oxford University Press, 2012.

Le Goff, J. *Una Edad Media en imágenes*, Barcelona, Paidós, 2009.

Lindberg, D.C., *The beginnings of Western science: The European scientific tradition in philosophical, religious, and institutional context, 600 B.C. to A.D. 1450*, Chicago, University of Chicago Press, 1992.

Llull, R. [1304] *Ascenso y descenso del entendimiento,* Barcelona, Folio, 2002.

Mâle, E. [1986] *El arte religioso del siglo XIII en Francia*, Madrid, Encuentro, 2001.

Molinos, M. de, [1675] *Manudictio Spiritualis*… Lipsiae, Waechtler, 1687.

Mujica Pinilla, R. «El arte y los sermones», en AA.VV, *El Barroco Peruano,* Lima, Banco de Crédito, pp. 218-313, 2002.

Rubin, M, *Emotion and Devotion*. *The Meaning of Mary in Medieval Religious Cultures.* Budapest: Central European University Press, 2009.

San Ildefonso, A. de, *Theologia mystica, sciencia y sabiduria de dios misteriosa.* Sin ciudad ni editor, 1683.

Savonarola, G. *De Simplicitate vite christiane. Aureus liber…,* Venecia, Lazaro de Soardis, 1504.

Schenone, H. *Santa María*, Buenos Aires, Educa, 2008.

Siracusano, G. (ed.) *La paleta del espanto, color y cultura en los cielos e infiernos en la pintura colonial andina*, Buenos Aires, UNSAM, 2010.

Valeri, V. *Opera ea quae ad adinventam ab ipso artem universalem scientiarum*, Argentorati [Estrasburgo], Lazari Zetzneri, 1617.

Vanegas Carrasco, C. *Trabajo final Historia del arte argentino y latinoamericano I*, IDAES-UnSAM, manuscrito inédito, 2010.

Vorágine, J. de, *Legenda aurea sanctorum,* Madrid, Juan García Infanzón, 1688.

Warburg, A., *Atlas Mnemosyne*, ed. M. Warnke, Madrid, Akal, 2012

Yates, F. A. [1982] *Ensayos reunidos*. Tomo I, México, Fondo de Cultura Económica, 1996.